\documentclass[twocolumn,aps,pra,showpacs,longbibliography,superscriptaddress]{revtex4-1}
\usepackage{amssymb,latexsym}
\usepackage{float}
\usepackage{mathrsfs}
\usepackage{lipsum}
\usepackage{graphicx}
\usepackage[caption=false]{subfig}
\usepackage{mathtools}
\usepackage{epstopdf}
\usepackage{xcolor}
\usepackage[colorlinks, linkcolor=red, anchorcolor=green, citecolor=green, urlcolor=blue]{hyperref}
\DeclareGraphicsExtensions{.pdf,.jpg,.png,.eps}
\usepackage[toc,page,title,titletoc,header]{appendix}
\usepackage{etoolbox}
\usepackage{breqn}

\makeatletter
\let\cat@comma@active\@empty
\makeatother

\begin{document}
	
\title{Performance advantage of quantum hypothesis testing for partially coherent optical sources}

\author{Jian-Dong Zhang}
\email[]{zhangjiandong@jsut.edu.cn}
\affiliation{School of Mathematics and Physics, Jiangsu University of Technology, Changzhou 213001, China}
\author{Kexin Zhang}
\affiliation{School of Mathematics and Physics, Jiangsu University of Technology, Changzhou 213001, China}
\author{Lili Hou}
\affiliation{School of Mathematics and Physics, Jiangsu University of Technology, Changzhou 213001, China}
\author{Shuai Wang}
\affiliation{School of Mathematics and Physics, Jiangsu University of Technology, Changzhou 213001, China}

\date{\today}
	
\begin{abstract}
Determining the presence of a potential optical source in the interest region is important for an imaging system and can be achieved by using hypothesis testing.
The previous studies assume that the potential source is completely incoherent. 
In this paper, this problem is generalized to the scenario with partially coherent sources and any prior probabilities. 
We compare the error probability limit given by the quantum Helstrom bound with the error probability given by direct decision based on the prior probability.
On this basis, the quantum-optimal detection advantage and detection-useless region are analyzed.
For practical purposes, we propose a specific detection strategy using binary spatial-mode demultiplexing, which can be used in the scenarios without any prior information.
This strategy shows superior detection performance and the results hold prospects for achieving super-resolved microscopic and astronomical imaging.
\end{abstract}

\maketitle

\section{Introduction}
The ability to distinguish one-versus-two incoherent point sources is one of the important metrics for evaluating the resolution of an optical imaging  system. 
This task becomes difficult when the two sources severely overlap.
In this regard, Rayleigh's criterion comes across as the most famous theory. 
The results indicate that two incoherent point sources cannot be distinguished when the separation between them is less than a diffraction-limited spatial size. 
In other words, the precision of estimating the separation between the two sources rapidly deteriorates as the separation decreases. 
This idea, although rough, had a profound impact for over a century.
Realizing sub-Rayleigh super-resolution imaging has always been one of the pursuit goals in many related fields.

With the continuous development of quantum theory, the use of quantum parameter estimation to analyze incoherent imaging problem has aroused a lot of interests. 
In 2016, Tsang \emph{et al.}  used the classical Fisher information to calculate the estimation precision for direct imaging, and the results were a manifestation of the Rayleigh's criterion \cite{PhysRevX.6.031033}. 
They provided the precision limit for separation estimation by using the quantum Fisher information. 
Surprisingly, the precision limit was independent of the actual separation, meaning that appropriate quantum detection can provide the estimation precision over direct imaging for a small separation.
Subsequently, some special detection strategies were proposed, such as super localization by image inversion interferometry (SLIVER) \cite{Nair:16,Schodt:23,PhysRevLett.117.190801}, spatial-mode demultiplexing (SPADE) \cite{PhysRevA.97.023830,PhysRevA.95.063847,Tsang_2017,Tan:23}, super-resolved position localization by inversion of coherence along an edge (SPLICE) \cite{PhysRevLett.118.070801}.
Given that two sources are generally not completely incoherent, the estimation of the separation between two partially incoherent sources has received a lot of attention \cite{Larson:18,Tsang:19,Larson:19}.
In addition, the impacts of some realistic factors have been analyzed, including misalignment, cross talk, and detector noise \cite{PhysRevLett.127.123604,Schlichtholz:24,PhysRevA.101.022323,PhysRevLett.126.120502,doi:10.1142/S0219749919410156,PhysRevLett.125.100501,Linowski_2023}. 
These beneficial explorations may useful for many applications like microscopic imaging and astronomical observations.

For a practical imaging task, prior to estimating the separation, it is generally necessary to ensure that there are two sources on the object plane and not just one.
To do this, one straightforward method would be to use hypothesis testing to determine whether a secondary source is present or not.
Related to this, many studies have focused on this subject over the past years \cite{PhysRevA.103.022406,zanforlin2022optical,PhysRevLett.127.130502,lu2018quantum}.
However, these relevant studies concentrated on the scenario with two  incoherent sources.
In this paper, we consider  a more general scenario where the two sources are partially coherent and prior probabilities of two hypotheses are arbitrary. 
We calculate the error probability limit given by the quantum Helstrom bound \cite{1055052}.
As a comparison, the error probability of direct decision based on prior probability is also analyzed.
For the most common scenarios without any prior information, we report a binary detection strategy which can provide superior performance advantage.

The remainder of this paper is organized as follows.
In Sec. \ref{s2}, we introduce our model and the fundamental principle.
Section \ref{s3} analyzes the quantum Helstrom bound and calculates the quantum-optimal detection advantage.
Section \ref{s4} gives a specific detection strategy based on binary SPADE, and the corresponding error probability is calculated.
Finally, we summarize our main results in Sec. \ref{s5}.

\section{Fundamental principle}
\label{s2}
\begin{figure*}[htbp]
	\centering
	\includegraphics[width=0.45\textwidth]{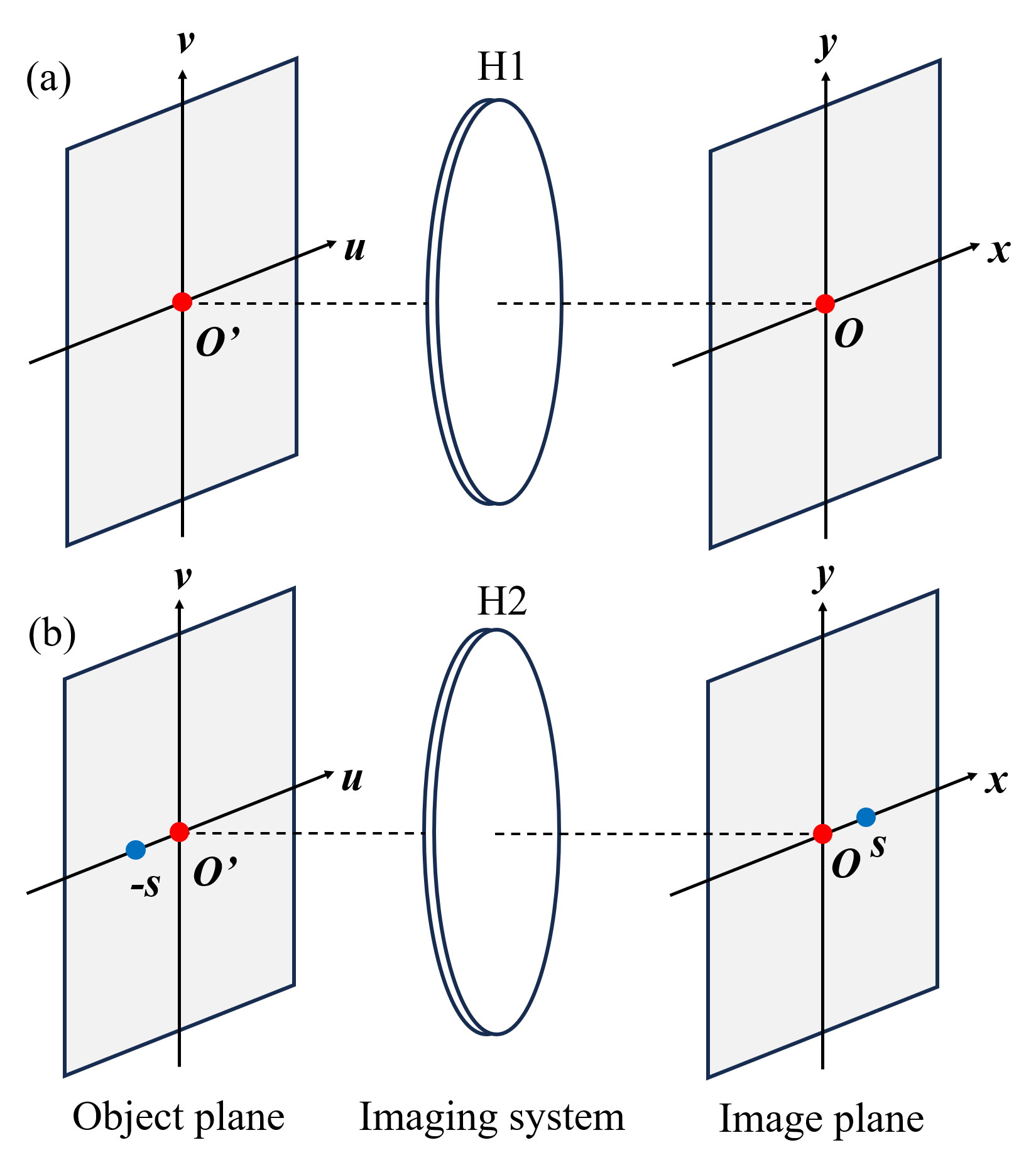}
	\caption{Schematic of the imaging by a spatially-invariant imaging system, (a) a single source; (b) two partially incoherent point sources.}
	\label{system}
\end{figure*}

In Fig. \ref{system}, we show the schematic of two hypotheses in our model.
A spatially-invariant imaging system images a single source or two partially incoherent sources on the image plane.
The primary source is positioned at 0, and the separation between it and the secondary source, if any, is positioned at $s$.
The task is to detect the photons on the image plane and to make a decision between the hypothesis $H_1$ (one source) and hypothesis $H_2$ (two sources).

For many scenarios in microscopic and astronomical imaging, the observable brightness is very faint.
In this regime, the quantum state on the image plane can be approximated as the superposition of zero and one photons, i.e.,
\begin{align}
	\rho = (1-\epsilon)\left| {\rm{zero}} \right\rangle \left\langle {\rm{zero}} \right| + \epsilon \left| {\rm{one}} \right\rangle \left\langle {\rm{one}} \right|
	\label{e0}
\end{align}
with $\epsilon \ll 1$ being the mean photon number, where $\left| {\rm{zero}} \right\rangle $ and $\left| {\rm{one}} \right\rangle $ stand for zero-photon and one-photon states.
Since zero-photon states provide no information, in what follows we merely consider the detection results originating from one-photon states.

Throughout this paper, we use ket vectors $\left| {\psi_0} \right\rangle$ and $\left| {\psi_s} \right\rangle$ to represent the states of photons emitted by two sources.
Without loss of generality, we assume that the point-spread function of the imaging system is Gaussian.
At this time, two ket vectors can be written as
\begin{align}
	\left| {\psi_0} \right\rangle = \int_{ - \infty }^\infty  {{\left( {\frac{1}{{2\pi {\sigma ^2}}}} \right)^{{1 \mathord{\left/
						{\vphantom {1 4}} \right.
						\kern-\nulldelimiterspace} 4}}}\exp \left( { - \frac{{{x^2}}}{{4{\sigma ^2}}}} \right) {\left| {x} \right\rangle}dx} 
	\label{}
\end{align}

\begin{align}
	\left| {\psi_s} \right\rangle = \int_{ - \infty }^\infty  {{\left( {\frac{1}{{2\pi {\sigma ^2}}}} \right)^{{1 \mathord{\left/
						{\vphantom {1 4}} \right.
						\kern-\nulldelimiterspace} 4}}}\exp \left[ { - \frac{{{(x-s)^2}}}{{4{\sigma ^2}}}} \right] {\left| {x} \right\rangle}dx} 
	\label{}
\end{align}
where $\sigma$ is the standard deviation. 
Further, it can be found that these two ket vectors are not orthogonal since
\begin{equation}
	\delta  = \left\langle {{{\psi _s}}}
	{\left | {\vphantom {{{\psi _s}} {{\psi _0}}}}
		\right. \kern-\nulldelimiterspace}
	{{{\psi _0}}} \right\rangle  = \left\langle {{{\psi _0}}}
	{\left | {\vphantom {{{\psi _0}} {{\psi _s}}}}
		\right. \kern-\nulldelimiterspace}
	{{{\psi _s}}} \right\rangle  = \exp \left( { - \frac{{{k^2}}}{8}} \right)
\end{equation}
with dimensionless separation $k \equiv {s}/{\sigma }$.
For two hypotheses $H_1$ and $H_2$, the density matrices on the image plane are found to be 

\begin{equation}
	{\rho _1} = \left| {{\psi _0}} \right\rangle \left\langle {{\psi _0}} \right|
\end{equation}

\begin{align}
	{\rho _2} = N\left[ {\left| {{\psi _0}} \right\rangle \left\langle {{\psi _0}} \right| + \left| {{\psi _s}} \right\rangle \left\langle {{\psi _s}} \right| +   \gamma  \left( {\left| {{\psi _0}} \right\rangle \left\langle {{\psi _s}} \right| + \left| {{\psi _s}} \right\rangle \left\langle {{\psi _0}} \right|} \right)} \right]
\end{align}
with normalized factor
\begin{equation}
	N = \frac{1}{{2\left( {1 + \delta \gamma \cos\theta } \right)}},
\end{equation}
where $\gamma$ and $\theta$ are the strength and phase of complex coherent degree.

\section{Quantum-optimal detection advantage}
\label{s3}

According to the quantum information theory, the minimal error probability over all possible detection strategies in hypothesis testing is govern by the the quantum Helstrom bound, which is defined as
\begin{align}
	{O_{\rm{err}}} = \min \left\{ {{P_{\rm{err}}}} \right\} = \frac{1}{2}\left( {1 - \left\| \Lambda \right\|} \right) 
	\label{}
\end{align}
with $\Lambda  = p{\rho _2} - \left( {1 - p} \right){\rho _1}$, where $p$ is the prior probability and $\left\| \cdot  \right\|$ denotes the trace norm of the matrix.
Especially, the trace norm equals to the sum of absolute values of the eigenvalues as the matrix $\Lambda$ is Hermitian  \cite{PhysRevA.71.062340}.
On the other hand, direct decision is a simple strategy at low cost.
Specifically, one can bypass detection and make a decision directly based on prior probabilities.
At this time, the error probability is given by
\begin{align}
	{{D_{\rm{err}}}} = \min \left\{ p,1-p \right\}. 
	\label{}
\end{align}

Here we define the ratio of these two error probabilities as quantum-optimal detection advantage, i.e.,
\begin{align}
	{{\cal A}_{\rm QOD}} = \frac{{D_{\rm{err}}}} { {{O_{\rm{err}}}} }. 
	\label{}
\end{align}
This value is always not less than 1 (${{\cal A}_{\rm QOD}} \ge 1$).
For some parameters, one may have ${{\cal A}_{\rm QOD}} > 1$, indicating that there exists at least an appropriate detection strategy providing superior performance beyond direct decision.
In particular, we can get ${{\cal A}_{\rm QOD}} = 1$ for some parameters, which suggests that direct decision is the optimal detection strategy, and the corresponding region is called the detection-useless region.

In order to calculate the quantum 
Helstrom bound, we reconstruct two orthogonal ket vectors through the use of Schmidt orthogonalization.
The specific ket vectors are as follows
\begin{equation}
	\left| 0 \right\rangle  = \left| {{\psi _0}} \right\rangle 
\end{equation}
\begin{equation}
	\left| 1 \right\rangle  = \frac{1}{{\sqrt {1 - {\delta ^2}} }}\left( {\left| {{\psi _s}} \right\rangle  - \delta \left| 0 \right\rangle } \right).
\end{equation}

On this basis, the density matrices can be rewritten as
\begin{equation}
	{\rho _1} = \left[ {\begin{array}{*{20}{c}}
			1&0\\
			0&0
	\end{array}} \right]
\end{equation}
\begin{align}
	{\rho _2} = N\left[ {\begin{array}{*{20}{c}}
			{1 + {\delta ^2} + 2\delta \gamma \cos \theta }&{\left( {\delta  + \gamma \cos \theta } \right)\sqrt {1 - {\delta ^2}} }\\
			{\left( {\delta  + \gamma \cos \theta } \right)\sqrt {1 - {\delta ^2}} }&{1 - {\delta ^2}}
	\end{array}} \right].
\end{align}

Finally, we get
\begin{equation}
	\Lambda  = N\left[ {\begin{array}{*{20}{c}}
			{p\left( {1 + {\delta ^2} + 2\delta \gamma \cos \theta } \right) + p - 1}&{p\left( {\delta  + \gamma \cos \theta } \right)\sqrt {1 - {\delta ^2}} }\\
			{p\left( {\delta  + \gamma \cos \theta } \right)\sqrt {1 - {\delta ^2}} }&{p\left( {1 - {\delta ^2}} \right)}
	\end{array}} \right].
\end{equation}
It is not difficult to find that we arrive at detection-useless region (${{\cal A}_{\rm QOD}} = 1$) when all eigenvalues of $\Lambda$ are positive or negative. 
As a result, we can give analytical expression for this region using sequential principal minor.

According to the theory of linear algebra, all eigenvalues of $\Lambda$ are positive if the matrix elements satisfy
\begin{equation}
	{\Lambda _{11}} > 0{\kern 5pt} \& {\kern 5pt}  {\Lambda _{11}}{\Lambda _{22}} - {\Lambda _{12}}{\Lambda _{21}} > 0
\end{equation}
The solutions for these two inequalities are given by  
\begin{equation}
	p > \frac{{2 + 2\delta \gamma \cos \theta }}{{3 + 4\delta \gamma \cos \theta  + {\delta ^2}}},
\end{equation}
\begin{equation}
	p > \frac{{2 + 2\delta \gamma \cos \theta }}{{3 + 2\delta \gamma \cos \theta  - {\gamma ^2}{{\cos }^2}\theta }}.
\end{equation}
It is worth noting that
\begin{equation}
	3 + 4\delta \gamma \cos \theta  + {\delta ^2} \ge 3 + 2\delta \gamma \cos \theta  - {\gamma ^2}{\cos ^2}\theta. 
\end{equation}
As a consequence, the detection-useless region is found to be
\begin{equation}
	p > \frac{{2 + 2\delta \gamma \cos \theta }}{{3 + 2\delta \gamma \cos \theta  - {\gamma ^2}{{\cos }^2}\theta }}.
\end{equation}
Similarly, all eigenvalues of $\Lambda$ are negative if we have
\begin{equation}
	{\Lambda _{11}} < 0{\kern 5pt} \&  {\kern 5pt} {\Lambda _{11}}{\Lambda _{22}} - {\Lambda _{12}}{\Lambda _{21}} > 0.
\end{equation}
It turns out that there is no solution for these two inequalities.

\begin{figure*}[htbp]
	\centering
	\includegraphics[width=0.35\textwidth]{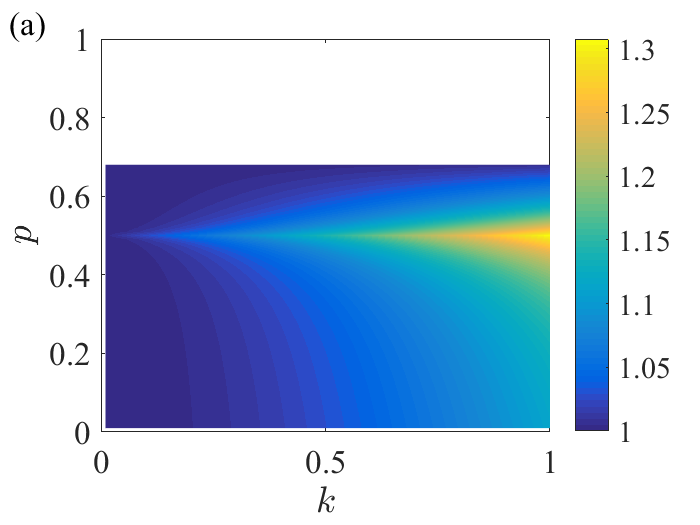}
	\includegraphics[width=0.35\textwidth]{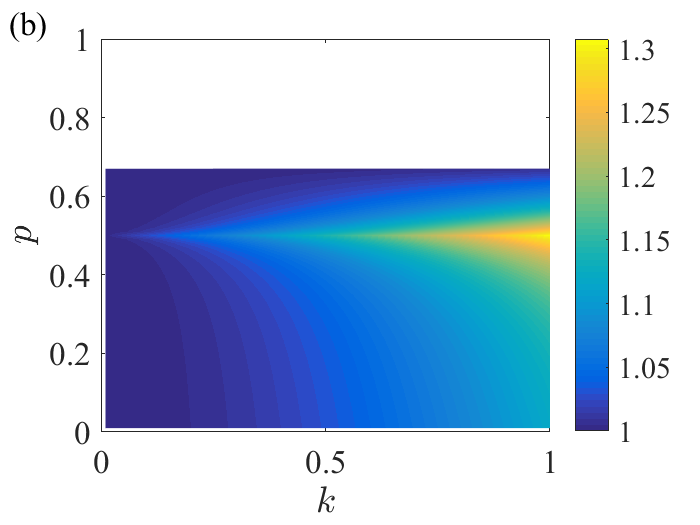}
	\includegraphics[width=0.35\textwidth]{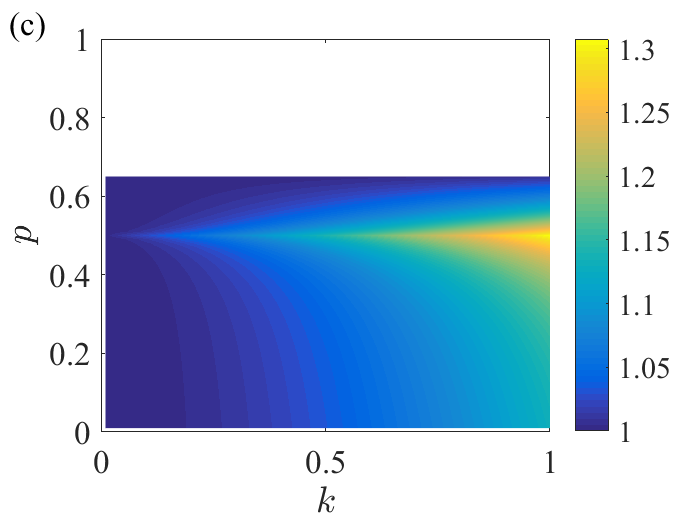}
	\includegraphics[width=0.35\textwidth]{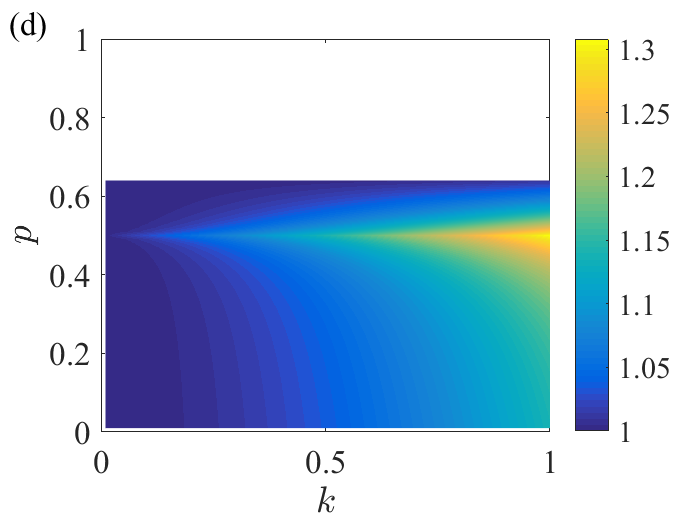}
	\caption{The quantum-optimal detection advantage ${\cal A}_{\rm QOD}$ versus the prior probability $p$ and separation $k$ with weak coherent degree $\gamma = 0.1$ and (a) $\theta= 0$, (b) $\theta= \pi/3$, (c) $\theta= 2\pi/3$, (d) $\theta = \pi$; blank: useless detection region (${\cal A}_{\rm QOD} = 1$).}
	\label{f2}
\end{figure*}

Based on the above results, one can analyze the quantum-optimal detection advantage.
In Fig. \ref{f2}, we show the result against the prior probability and separation with weak coherent degree.
It can be seen that the impacts of phase on the quantum-optimal detection advantage is not significant.
The increase in phase results in a slight extension of the  detection-useless region. 
At this point, the impacts of coherent degree on hypothesis testing mainly come from strength.
The quantum-optimal detection advantage increases with the increase of separation. 
This is due to the fact that two sources with a large separation provide more information and can be easily distinguished.
In addition, the quantum-optimal detection advantage in general occurs at $p=0.5$.
This originates from the reason that direct decision has no prior information at this time.

\begin{figure*}[htbp]
	\centering
	\includegraphics[width=0.35\textwidth]{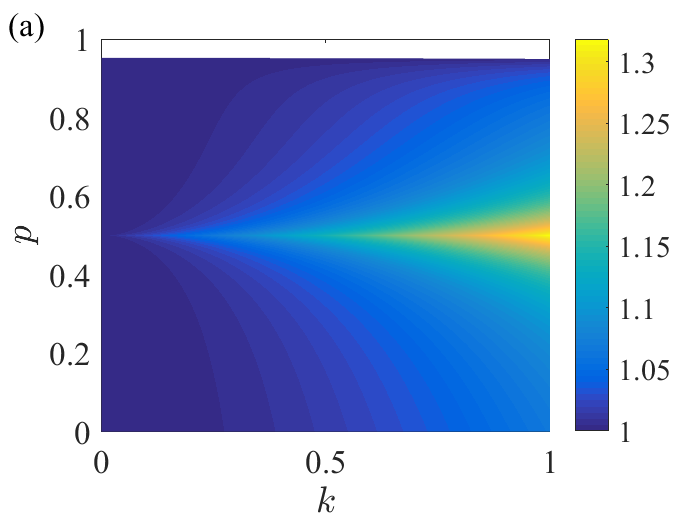}
	\includegraphics[width=0.35\textwidth]{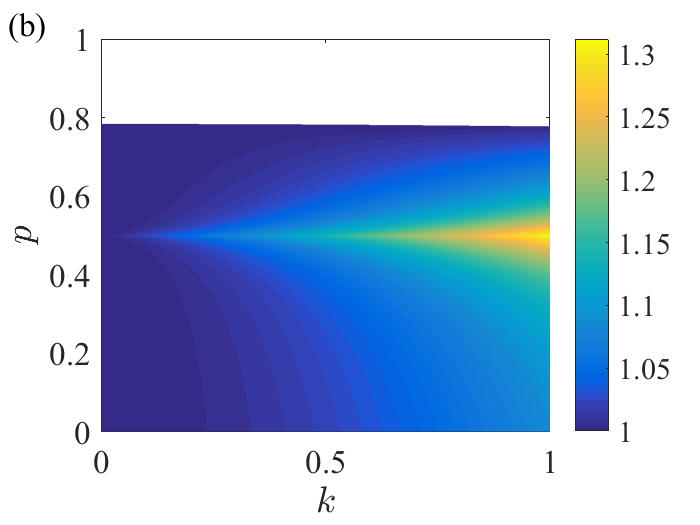}
	\includegraphics[width=0.35\textwidth]{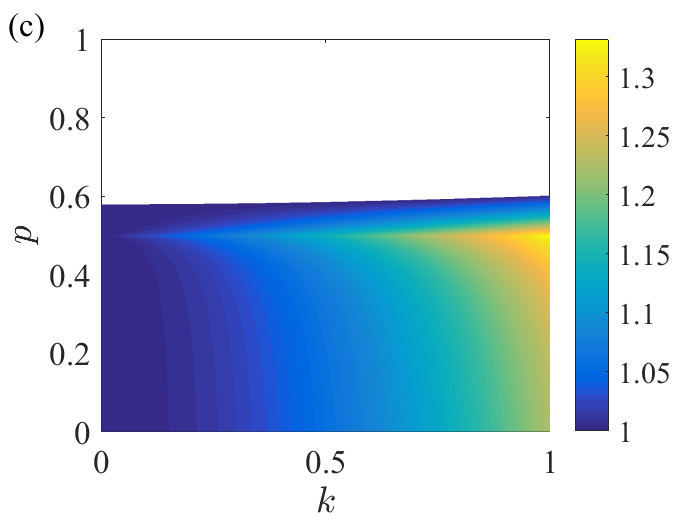}
	\includegraphics[width=0.35\textwidth]{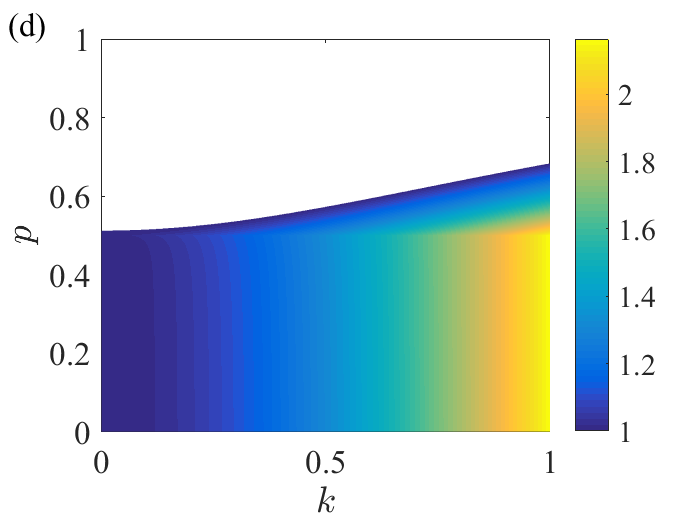}
	\caption{The quantum-optimal detection advantage ${\cal A}_{\rm QOD}$ versus the prior probability $p$ and separation $k$ with strong coherent degree $\gamma = 0.9$ and (a) $\theta= 0$, (b) $\theta= \pi/3$, (c) $\theta= 2\pi/3$, (d) $\theta = \pi$; blank: useless detection region (${\cal A}_{\rm QOD} = 1$).}
	\label{f3}
\end{figure*}

As a comparison, in Fig. \ref{f3} we give the quantum-optimal detection advantage as a function of the prior probability and separation with strong coherent degree.
Many phenomena in the figure are similar to that in Fig. \ref{f2}.
Unlike the scenario with weak coherent degree, the detection-useless region changes significantly with the increase of phase.
This indicates that the phase between two sources can provide useful information for detection.
Interestingly, in-phase ($\theta = 0$) scenario can shrink useless detection region whereas out-phase ($\theta = \pi$) scenario can improve the quantum-optimal detection advantage.

\section{Detection performance with binary SPADE}
\label{s4}

In the previous section, we analyzed the quantum-optimal detection performance based on the quantum Helstrom bound.
Unfortunately, the optimal detection strategy that can approach this bound is often difficult to implement.
Here we report the detection performance of a binary SPADE strategy.

For a standard hypothesis testing, one needs to know the true values of all parameters in advance.
However, this is generally not achievable for many realistic scenarios. 
If we know the true value of the separation, the result can be determined without any detection since $k = 0$ and $k \neq 0$ indicate a single source and two partially coherent sources, respectively.
In a realistic scenario, we have no prior information ($p = 0.5$).
For this reason, we propose a binary detection strategy based on SPADE, which can work in this scenario.
Specifically, we classify the photons according to Gaussian and non-Gaussian modes, and the mode classifier is aligned with the center of the image plane.
Since the first source is known, the center of the image plane is easy to determine. 
On the other hand, Gaussian mode can be selected through a fiber or waveguide \cite{PhysRevA.101.063830,PhysRevA.95.063847}.
Hence, binary SPADE is not difficult to implement.

In experiments, we can deploy two detectors to record photons corresponding to Gaussian and non-Gaussian modes, respectively.
Then the probabilities of all detection events in $H_1$ hypothesis are given by
\begin{align}
	{\rm{Pr}}({\rm{On,Off}}) =&{\kern 2pt} 1, \\
	\label{}
	{\rm{Pr}}({\rm{On,On}}) = &{\kern 2pt} 0, \\
	\label{}
	{\rm{Pr}}({\rm{Off,Off}}) = &{\kern 2pt} 0, \\
	\label{}
	{\rm{Pr}}({\rm{Off,On}}) = &{\kern 2pt} 0, 
	\label{}
\end{align}
while those in $H_2$ hypothesis are given by
\begin{align}
	{\rm{Pr}}({\rm{On,Off}}) = & {\kern 2pt}\frac{{1 + {\delta ^2} + 2\delta \gamma \cos \theta }}{{2( 1 + \delta \gamma \cos \theta  )}}, \\
	\label{}
	{\rm{Pr}}({\rm{On,On}}) = & {\kern 2pt} 0, \\
	\label{}
	{\rm{Pr}}({\rm{Off,Off}}) = &{\kern 2pt} 0, \\
	\label{}
	{\rm{Pr}}({\rm{Off,On}}) = &{\kern 2pt} 1 - \frac{{1 + {\delta ^2} + 2\delta \gamma \cos \theta }}{{2( 1 + \delta \gamma \cos \theta  )}}, 
	\label{f4}
\end{align}

In terms of the above probabilities, the decision rule can be determined.
If a photon is detected by the second detector (On), then we accept $H_2$ hypothesis, otherwise (Off) we accept $H_1$ hypothesis.
Based on this rule, the error probability is found to be 
\begin{equation}
	{P_{\rm{err}}} = \frac{{1 + {\delta ^2} + 2\delta \gamma \cos \theta }}{{4( 1 + \delta \gamma \cos \theta  )}}. 
	\label{}
\end{equation}
The ratio of above result to the error probability of direct decision is defined as the detection advantage of binary SPADE, i.e.,
\begin{align}
	{{\cal A}_{\rm D}} = \frac{{D_{\rm{err}}}} { {{P_{\rm{err}}}} }. 
	\label{}
\end{align}

\begin{figure*}[htbp]
	\centering
	\includegraphics[width=0.42\textwidth]{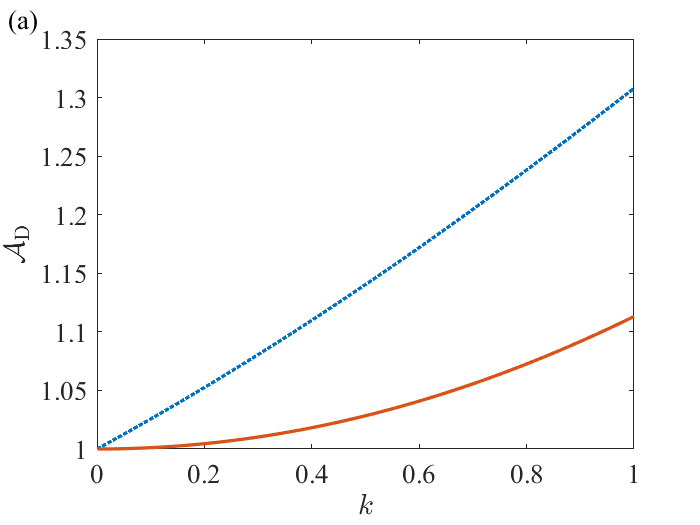}
	\includegraphics[width=0.42\textwidth]{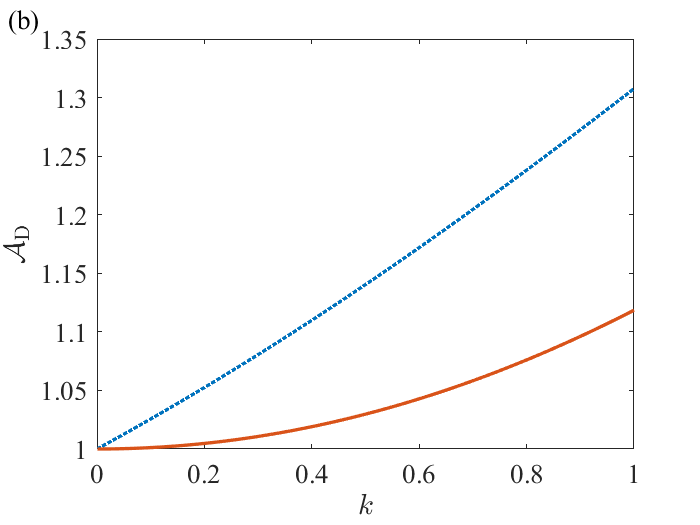}
	\includegraphics[width=0.42\textwidth]{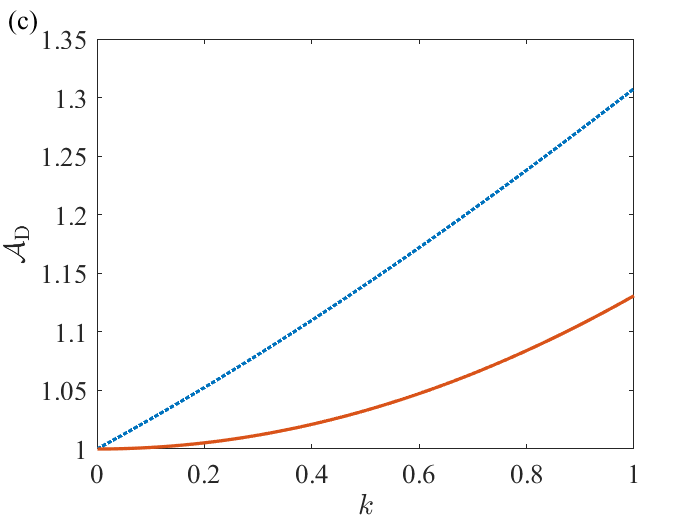}
	\includegraphics[width=0.42\textwidth]{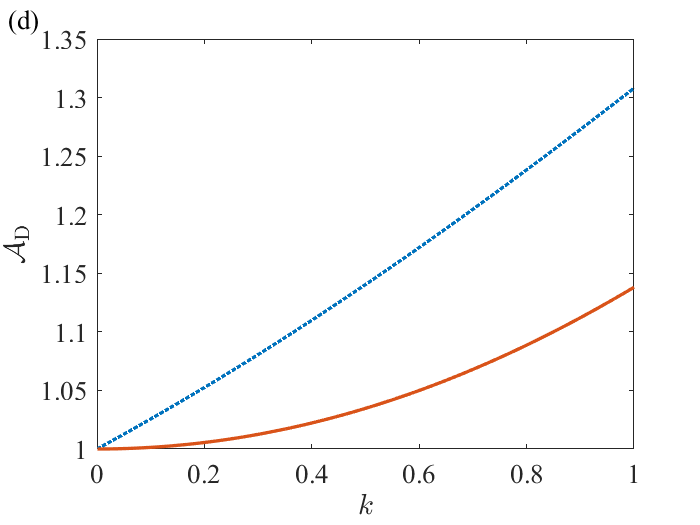}
	\caption{The quantum-optimal detection advantage $A_{\rm QOD}$ (blue line) and detection advantage $A_{\rm D}$ (red line) versus  separation $k$ with weak coherent degree $\gamma = 0.9$ and (a) $\theta= 0$; (b) $\theta= \pi/3$; (c) $\theta= 2\pi/3$; (d) $\theta = \pi$.}
	\label{f4}
\end{figure*}

\begin{figure*}[htbp]
	\centering
	\includegraphics[width=0.42\textwidth]{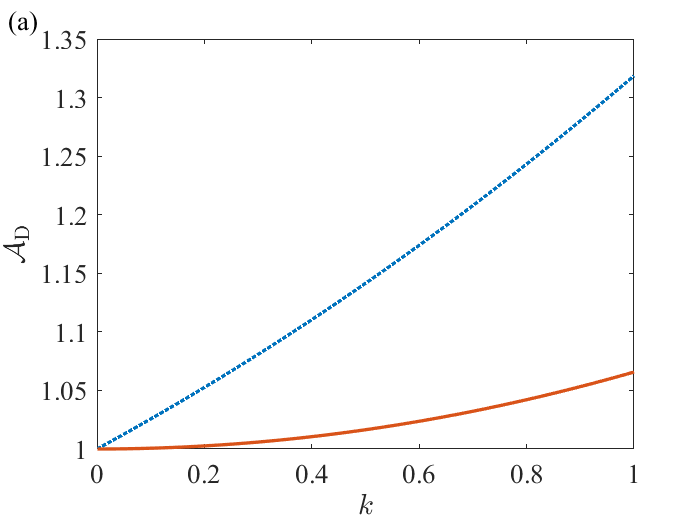}
	\includegraphics[width=0.42\textwidth]{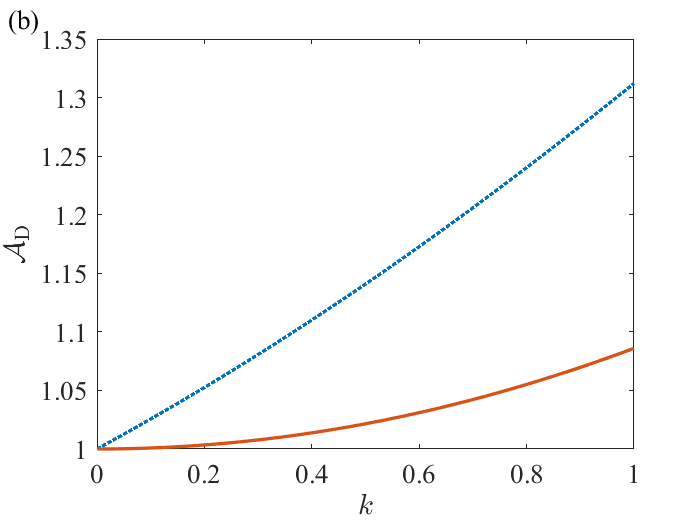}
	\includegraphics[width=0.42\textwidth]{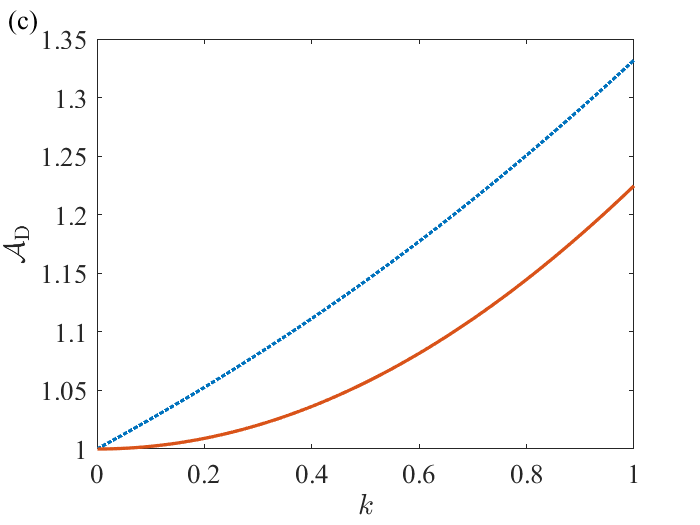}
	\includegraphics[width=0.42\textwidth]{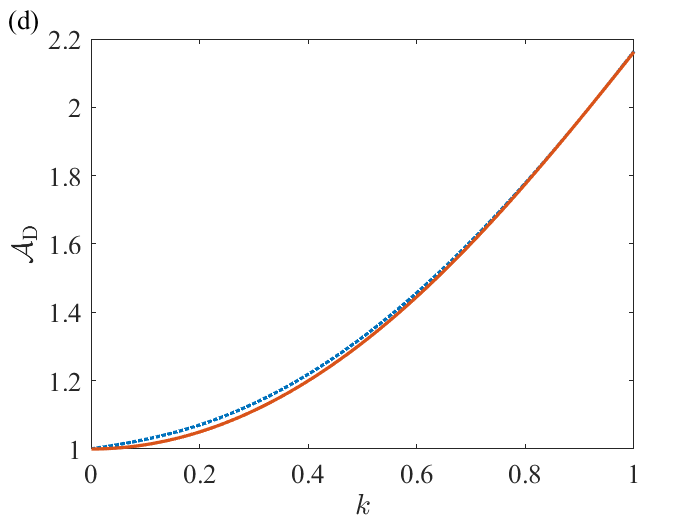}
	\caption{The quantum-optimal detection advantage $A_{\rm QOD}$ (blue line) and detection advantage $A_{\rm D}$ (red line) versus  separation $k$ with strong coherent degree $\gamma = 0.9$ and (a) $\theta= 0$; (b) $\theta= \pi/3$; (c) $\theta= 2\pi/3$; (d) $\theta = \pi$.}
	\label{f5}
\end{figure*}

To evaluate the detection performance of binary SPADE, Figs. \ref{f4} and \ref{f5} provide the quantum-optimal detection advantage as well as detection advantage given by binary SPADE with weak and strong coherent degrees. 
It turns out that, for weak coherent degree, the quantum-optimal detection advantage is almost constant regardless the the phase, while the detection advantage of binary SPADE slowly improves as the phase increases. 
The detection advantage of binary SPADE is about half of the quantum-optimal detection advantage, indicating that it is a sub-optimal detection strategy regarding weak coherent degree.

For strong coherent degree, both the quantum-optimal detection advantage and the detection advantage of binary SPADE are improved with the increase of the phase. 
In contrast, the detection advantage of binary SPADE has a faster rate.
In particular, for two sources with out-phase, the detection advantage of binary SPADE approaches the quantum-optimal detection advantage. 
This suggests that binary SPADE is the optimal detection strategy for strong coherent degree and out-phase.

\section{Conclusions}
\label{s5}
In summary, we studied the quantum hypothesis testing with partially coherent sources and any prior probability.
We calculated the quantum Helstrom bound and compared it with the error probability of direct decision based on the prior probability.
The quantum-optimal detection advantage and detection-useless region were discussed with weak and strong coherent degrees.
Based on spatial-mode demultiplexing, we proposed a binary detection strategy which can work in the scenario without any prior information.
Superior detection performance was demonstrated and may be useful for practical applications such as microscopic and astronomical imaging.

\section*{Acknowledgment} 
This work was supported by the Program of Zhongwu Young Innovative Talents of Jiangsu University of Technology (20230013).

%\bibliography{sample.bib}

\begin{thebibliography}{26}%
\makeatletter
\providecommand \@ifxundefined [1]{%
 \@ifx{#1\undefined}
}%
\providecommand \@ifnum [1]{%
 \ifnum #1\expandafter \@firstoftwo
 \else \expandafter \@secondoftwo
 \fi
}%
\providecommand \@ifx [1]{%
 \ifx #1\expandafter \@firstoftwo
 \else \expandafter \@secondoftwo
 \fi
}%
\providecommand \natexlab [1]{#1}%
\providecommand \enquote  [1]{``#1''}%
\providecommand \bibnamefont  [1]{#1}%
\providecommand \bibfnamefont [1]{#1}%
\providecommand \citenamefont [1]{#1}%
\providecommand \href@noop [0]{\@secondoftwo}%
\providecommand \href [0]{\begingroup \@sanitize@url \@href}%
\providecommand \@href[1]{\@@startlink{#1}\@@href}%
\providecommand \@@href[1]{\endgroup#1\@@endlink}%
\providecommand \@sanitize@url [0]{\catcode `\\12\catcode `\$12\catcode `\&12\catcode `\#12\catcode `\^12\catcode `\_12\catcode `\%12\relax}%
\providecommand \@@startlink[1]{}%
\providecommand \@@endlink[0]{}%
\providecommand \url  [0]{\begingroup\@sanitize@url \@url }%
\providecommand \@url [1]{\endgroup\@href {#1}{\urlprefix }}%
\providecommand \urlprefix  [0]{URL }%
\providecommand \Eprint [0]{\href }%
\providecommand \doibase [0]{http://dx.doi.org/}%
\providecommand \selectlanguage [0]{\@gobble}%
\providecommand \bibinfo  [0]{\@secondoftwo}%
\providecommand \bibfield  [0]{\@secondoftwo}%
\providecommand \translation [1]{[#1]}%
\providecommand \BibitemOpen [0]{}%
\providecommand \bibitemStop [0]{}%
\providecommand \bibitemNoStop [0]{.\EOS\space}%
\providecommand \EOS [0]{\spacefactor3000\relax}%
\providecommand \BibitemShut  [1]{\csname bibitem#1\endcsname}%
\let\auto@bib@innerbib\@empty
%</preamble>
\bibitem [{\citenamefont {Tsang}\ \emph {et~al.}(2016)\citenamefont {Tsang}, \citenamefont {Nair},\ and\ \citenamefont {Lu}}]{PhysRevX.6.031033}%
  \BibitemOpen
  \bibfield  {author} {\bibinfo {author} {\bibfnamefont {Mankei}\ \bibnamefont {Tsang}}, \bibinfo {author} {\bibfnamefont {Ranjith}\ \bibnamefont {Nair}}, \ and\ \bibinfo {author} {\bibfnamefont {Xiao-Ming}\ \bibnamefont {Lu}},\ }\bibfield  {title} {\enquote {\bibinfo {title} {Quantum theory of superresolution for two incoherent optical point sources},}\ }\href {\doibase 10.1103/PhysRevX.6.031033} {\bibfield  {journal} {\bibinfo  {journal} {Phys. Rev. X}\ }\textbf {\bibinfo {volume} {6}},\ \bibinfo {pages} {031033} (\bibinfo {year} {2016})}\BibitemShut {NoStop}%
\bibitem [{\citenamefont {Nair}\ and\ \citenamefont {Tsang}(2016{\natexlab{a}})}]{Nair:16}%
  \BibitemOpen
  \bibfield  {author} {\bibinfo {author} {\bibfnamefont {Ranjith}\ \bibnamefont {Nair}}\ and\ \bibinfo {author} {\bibfnamefont {Mankei}\ \bibnamefont {Tsang}},\ }\bibfield  {title} {\enquote {\bibinfo {title} {Interferometric superlocalization of two incoherent optical point sources},}\ }\href {\doibase 10.1364/OE.24.003684} {\bibfield  {journal} {\bibinfo  {journal} {Opt. Express}\ }\textbf {\bibinfo {volume} {24}},\ \bibinfo {pages} {3684--3701} (\bibinfo {year} {2016}{\natexlab{a}})}\BibitemShut {NoStop}%
\bibitem [{\citenamefont {Schodt}\ \emph {et~al.}(2023)\citenamefont {Schodt}, \citenamefont {Cutler}, \citenamefont {Becerra},\ and\ \citenamefont {Lidke}}]{Schodt:23}%
  \BibitemOpen
  \bibfield  {author} {\bibinfo {author} {\bibfnamefont {David~J.}\ \bibnamefont {Schodt}}, \bibinfo {author} {\bibfnamefont {Patrick~J.}\ \bibnamefont {Cutler}}, \bibinfo {author} {\bibfnamefont {Francisco~E.}\ \bibnamefont {Becerra}}, \ and\ \bibinfo {author} {\bibfnamefont {Keith~A.}\ \bibnamefont {Lidke}},\ }\bibfield  {title} {\enquote {\bibinfo {title} {Tolerance to aberration and misalignment in a two-point-resolving image inversion interferometer},}\ }\href {\doibase 10.1364/OE.487808} {\bibfield  {journal} {\bibinfo  {journal} {Opt. Express}\ }\textbf {\bibinfo {volume} {31}},\ \bibinfo {pages} {16393--16405} (\bibinfo {year} {2023})}\BibitemShut {NoStop}%
\bibitem [{\citenamefont {Nair}\ and\ \citenamefont {Tsang}(2016{\natexlab{b}})}]{PhysRevLett.117.190801}%
  \BibitemOpen
  \bibfield  {author} {\bibinfo {author} {\bibfnamefont {Ranjith}\ \bibnamefont {Nair}}\ and\ \bibinfo {author} {\bibfnamefont {Mankei}\ \bibnamefont {Tsang}},\ }\bibfield  {title} {\enquote {\bibinfo {title} {Far-field superresolution of thermal electromagnetic sources at the quantum limit},}\ }\href {\doibase 10.1103/PhysRevLett.117.190801} {\bibfield  {journal} {\bibinfo  {journal} {Phys. Rev. Lett.}\ }\textbf {\bibinfo {volume} {117}},\ \bibinfo {pages} {190801} (\bibinfo {year} {2016}{\natexlab{b}})}\BibitemShut {NoStop}%
\bibitem [{\citenamefont {Tsang}(2018)}]{PhysRevA.97.023830}%
  \BibitemOpen
  \bibfield  {author} {\bibinfo {author} {\bibfnamefont {Mankei}\ \bibnamefont {Tsang}},\ }\bibfield  {title} {\enquote {\bibinfo {title} {Subdiffraction incoherent optical imaging via spatial-mode demultiplexing: Semiclassical treatment},}\ }\href {\doibase 10.1103/PhysRevA.97.023830} {\bibfield  {journal} {\bibinfo  {journal} {Phys. Rev. A}\ }\textbf {\bibinfo {volume} {97}},\ \bibinfo {pages} {023830} (\bibinfo {year} {2018})}\BibitemShut {NoStop}%
\bibitem [{\citenamefont {Ang}\ \emph {et~al.}(2017)\citenamefont {Ang}, \citenamefont {Nair},\ and\ \citenamefont {Tsang}}]{PhysRevA.95.063847}%
  \BibitemOpen
  \bibfield  {author} {\bibinfo {author} {\bibfnamefont {Shan~Zheng}\ \bibnamefont {Ang}}, \bibinfo {author} {\bibfnamefont {Ranjith}\ \bibnamefont {Nair}}, \ and\ \bibinfo {author} {\bibfnamefont {Mankei}\ \bibnamefont {Tsang}},\ }\bibfield  {title} {\enquote {\bibinfo {title} {Quantum limit for two-dimensional resolution of two incoherent optical point sources},}\ }\href {\doibase 10.1103/PhysRevA.95.063847} {\bibfield  {journal} {\bibinfo  {journal} {Phys. Rev. A}\ }\textbf {\bibinfo {volume} {95}},\ \bibinfo {pages} {063847} (\bibinfo {year} {2017})}\BibitemShut {NoStop}%
\bibitem [{\citenamefont {Tsang}(2017)}]{Tsang_2017}%
  \BibitemOpen
  \bibfield  {author} {\bibinfo {author} {\bibfnamefont {Mankei}\ \bibnamefont {Tsang}},\ }\bibfield  {title} {\enquote {\bibinfo {title} {Subdiffraction incoherent optical imaging via spatial-mode demultiplexing},}\ }\href {\doibase 10.1088/1367-2630/aa60ee} {\bibfield  {journal} {\bibinfo  {journal} {New J. Phys.}\ }\textbf {\bibinfo {volume} {19}},\ \bibinfo {pages} {023054} (\bibinfo {year} {2017})}\BibitemShut {NoStop}%
\bibitem [{\citenamefont {Tan}\ \emph {et~al.}(2023)\citenamefont {Tan}, \citenamefont {Qi}, \citenamefont {Chen}, \citenamefont {Danner}, \citenamefont {Kanchanawong},\ and\ \citenamefont {Tsang}}]{Tan:23}%
  \BibitemOpen
  \bibfield  {author} {\bibinfo {author} {\bibfnamefont {Xiao-Jie}\ \bibnamefont {Tan}}, \bibinfo {author} {\bibfnamefont {Luo}\ \bibnamefont {Qi}}, \bibinfo {author} {\bibfnamefont {Lianwei}\ \bibnamefont {Chen}}, \bibinfo {author} {\bibfnamefont {Aaron~J.}\ \bibnamefont {Danner}}, \bibinfo {author} {\bibfnamefont {Pakorn}\ \bibnamefont {Kanchanawong}}, \ and\ \bibinfo {author} {\bibfnamefont {Mankei}\ \bibnamefont {Tsang}},\ }\bibfield  {title} {\enquote {\bibinfo {title} {Quantum-inspired superresolution for incoherent imaging},}\ }\href {\doibase 10.1364/OPTICA.493227} {\bibfield  {journal} {\bibinfo  {journal} {Optica}\ }\textbf {\bibinfo {volume} {10}},\ \bibinfo {pages} {1189--1194} (\bibinfo {year} {2023})}\BibitemShut {NoStop}%
\bibitem [{\citenamefont {Tham}\ \emph {et~al.}(2017)\citenamefont {Tham}, \citenamefont {Ferretti},\ and\ \citenamefont {Steinberg}}]{PhysRevLett.118.070801}%
  \BibitemOpen
  \bibfield  {author} {\bibinfo {author} {\bibfnamefont {Weng-Kian}\ \bibnamefont {Tham}}, \bibinfo {author} {\bibfnamefont {Hugo}\ \bibnamefont {Ferretti}}, \ and\ \bibinfo {author} {\bibfnamefont {Aephraim~M.}\ \bibnamefont {Steinberg}},\ }\bibfield  {title} {\enquote {\bibinfo {title} {Beating rayleigh's curse by imaging using phase information},}\ }\href {\doibase 10.1103/PhysRevLett.118.070801} {\bibfield  {journal} {\bibinfo  {journal} {Phys. Rev. Lett.}\ }\textbf {\bibinfo {volume} {118}},\ \bibinfo {pages} {070801} (\bibinfo {year} {2017})}\BibitemShut {NoStop}%
\bibitem [{\citenamefont {Larson}\ and\ \citenamefont {Saleh}(2018)}]{Larson:18}%
  \BibitemOpen
  \bibfield  {author} {\bibinfo {author} {\bibfnamefont {Walker}\ \bibnamefont {Larson}}\ and\ \bibinfo {author} {\bibfnamefont {Bahaa E.~A.}\ \bibnamefont {Saleh}},\ }\bibfield  {title} {\enquote {\bibinfo {title} {Resurgence of rayleigh's curse in the presence of partial coherence},}\ }\href {\doibase 10.1364/OPTICA.5.001382} {\bibfield  {journal} {\bibinfo  {journal} {Optica}\ }\textbf {\bibinfo {volume} {5}},\ \bibinfo {pages} {1382--1389} (\bibinfo {year} {2018})}\BibitemShut {NoStop}%
\bibitem [{\citenamefont {Tsang}\ and\ \citenamefont {Nair}(2019)}]{Tsang:19}%
  \BibitemOpen
  \bibfield  {author} {\bibinfo {author} {\bibfnamefont {Mankei}\ \bibnamefont {Tsang}}\ and\ \bibinfo {author} {\bibfnamefont {Ranjith}\ \bibnamefont {Nair}},\ }\bibfield  {title} {\enquote {\bibinfo {title} {Resurgence of rayleigh's curse in the presence of partial coherence: comment},}\ }\href {\doibase 10.1364/OPTICA.6.000400} {\bibfield  {journal} {\bibinfo  {journal} {Optica}\ }\textbf {\bibinfo {volume} {6}},\ \bibinfo {pages} {400--401} (\bibinfo {year} {2019})}\BibitemShut {NoStop}%
\bibitem [{\citenamefont {Larson}\ and\ \citenamefont {Saleh}(2019)}]{Larson:19}%
  \BibitemOpen
  \bibfield  {author} {\bibinfo {author} {\bibfnamefont {Walker}\ \bibnamefont {Larson}}\ and\ \bibinfo {author} {\bibfnamefont {Bahaa E.~A.}\ \bibnamefont {Saleh}},\ }\bibfield  {title} {\enquote {\bibinfo {title} {Resurgence of rayleigh's curse in the presence of partial coherence: reply},}\ }\href {\doibase 10.1364/OPTICA.6.000402} {\bibfield  {journal} {\bibinfo  {journal} {Optica}\ }\textbf {\bibinfo {volume} {6}},\ \bibinfo {pages} {402--403} (\bibinfo {year} {2019})}\BibitemShut {NoStop}%
\bibitem [{\citenamefont {Sorelli}\ \emph {et~al.}(2021)\citenamefont {Sorelli}, \citenamefont {Gessner}, \citenamefont {Walschaers},\ and\ \citenamefont {Treps}}]{PhysRevLett.127.123604}%
  \BibitemOpen
  \bibfield  {author} {\bibinfo {author} {\bibfnamefont {Giacomo}\ \bibnamefont {Sorelli}}, \bibinfo {author} {\bibfnamefont {Manuel}\ \bibnamefont {Gessner}}, \bibinfo {author} {\bibfnamefont {Mattia}\ \bibnamefont {Walschaers}}, \ and\ \bibinfo {author} {\bibfnamefont {Nicolas}\ \bibnamefont {Treps}},\ }\bibfield  {title} {\enquote {\bibinfo {title} {Optimal observables and estimators for practical superresolution imaging},}\ }\href {\doibase 10.1103/PhysRevLett.127.123604} {\bibfield  {journal} {\bibinfo  {journal} {Phys. Rev. Lett.}\ }\textbf {\bibinfo {volume} {127}},\ \bibinfo {pages} {123604} (\bibinfo {year} {2021})}\BibitemShut {NoStop}%
\bibitem [{\citenamefont {Schlichtholz}\ \emph {et~al.}(2024)\citenamefont {Schlichtholz}, \citenamefont {Linowski}, \citenamefont {Walschaers}, \citenamefont {Treps}, \citenamefont {Rudnicki},\ and\ \citenamefont {Sorelli}}]{Schlichtholz:24}%
  \BibitemOpen
  \bibfield  {author} {\bibinfo {author} {\bibfnamefont {Konrad}\ \bibnamefont {Schlichtholz}}, \bibinfo {author} {\bibfnamefont {Tomasz}\ \bibnamefont {Linowski}}, \bibinfo {author} {\bibfnamefont {Mattia}\ \bibnamefont {Walschaers}}, \bibinfo {author} {\bibfnamefont {Nicolas}\ \bibnamefont {Treps}}, \bibinfo {author} {\bibfnamefont {{\L}ukasz}\ \bibnamefont {Rudnicki}}, \ and\ \bibinfo {author} {\bibfnamefont {Giacomo}\ \bibnamefont {Sorelli}},\ }\bibfield  {title} {\enquote {\bibinfo {title} {Practical tests for sub-rayleigh source discriminations with imperfect demultiplexers},}\ }\href {\doibase 10.1364/OPTICAQ.502459} {\bibfield  {journal} {\bibinfo  {journal} {Optica Quantum}\ }\textbf {\bibinfo {volume} {2}},\ \bibinfo {pages} {29--34} (\bibinfo {year} {2024})}\BibitemShut {NoStop}%
\bibitem [{\citenamefont {Lupo}(2020)}]{PhysRevA.101.022323}%
  \BibitemOpen
  \bibfield  {author} {\bibinfo {author} {\bibfnamefont {Cosmo}\ \bibnamefont {Lupo}},\ }\bibfield  {title} {\enquote {\bibinfo {title} {Subwavelength quantum imaging with noisy detectors},}\ }\href {\doibase 10.1103/PhysRevA.101.022323} {\bibfield  {journal} {\bibinfo  {journal} {Phys. Rev. A}\ }\textbf {\bibinfo {volume} {101}},\ \bibinfo {pages} {022323} (\bibinfo {year} {2020})}\BibitemShut {NoStop}%
\bibitem [{\citenamefont {Oh}\ \emph {et~al.}(2021)\citenamefont {Oh}, \citenamefont {Zhou}, \citenamefont {Wong},\ and\ \citenamefont {Jiang}}]{PhysRevLett.126.120502}%
  \BibitemOpen
  \bibfield  {author} {\bibinfo {author} {\bibfnamefont {Changhun}\ \bibnamefont {Oh}}, \bibinfo {author} {\bibfnamefont {Sisi}\ \bibnamefont {Zhou}}, \bibinfo {author} {\bibfnamefont {Yat}\ \bibnamefont {Wong}}, \ and\ \bibinfo {author} {\bibfnamefont {Liang}\ \bibnamefont {Jiang}},\ }\bibfield  {title} {\enquote {\bibinfo {title} {Quantum limits of superresolution in a noisy environment},}\ }\href {\doibase 10.1103/PhysRevLett.126.120502} {\bibfield  {journal} {\bibinfo  {journal} {Phys. Rev. Lett.}\ }\textbf {\bibinfo {volume} {126}},\ \bibinfo {pages} {120502} (\bibinfo {year} {2021})}\BibitemShut {NoStop}%
\bibitem [{\citenamefont {Len}\ \emph {et~al.}(2020)\citenamefont {Len}, \citenamefont {Datta}, \citenamefont {Parniak},\ and\ \citenamefont {Banaszek}}]{doi:10.1142/S0219749919410156}%
  \BibitemOpen
  \bibfield  {author} {\bibinfo {author} {\bibfnamefont {Yink~Loong}\ \bibnamefont {Len}}, \bibinfo {author} {\bibfnamefont {Chandan}\ \bibnamefont {Datta}}, \bibinfo {author} {\bibfnamefont {Micha\l{}}\ \bibnamefont {Parniak}}, \ and\ \bibinfo {author} {\bibfnamefont {Konrad}\ \bibnamefont {Banaszek}},\ }\bibfield  {title} {\enquote {\bibinfo {title} {Resolution limits of spatial mode demultiplexing with noisy detection},}\ }\href {\doibase 10.1142/S0219749919410156} {\bibfield  {journal} {\bibinfo  {journal} {Int. J. Quantum Inf.}\ }\textbf {\bibinfo {volume} {18}},\ \bibinfo {pages} {1941015} (\bibinfo {year} {2020})}\BibitemShut {NoStop}%
\bibitem [{\citenamefont {Gessner}\ \emph {et~al.}(2020)\citenamefont {Gessner}, \citenamefont {Fabre},\ and\ \citenamefont {Treps}}]{PhysRevLett.125.100501}%
  \BibitemOpen
  \bibfield  {author} {\bibinfo {author} {\bibfnamefont {Manuel}\ \bibnamefont {Gessner}}, \bibinfo {author} {\bibfnamefont {Claude}\ \bibnamefont {Fabre}}, \ and\ \bibinfo {author} {\bibfnamefont {Nicolas}\ \bibnamefont {Treps}},\ }\bibfield  {title} {\enquote {\bibinfo {title} {Superresolution limits from measurement crosstalk},}\ }\href {\doibase 10.1103/PhysRevLett.125.100501} {\bibfield  {journal} {\bibinfo  {journal} {Phys. Rev. Lett.}\ }\textbf {\bibinfo {volume} {125}},\ \bibinfo {pages} {100501} (\bibinfo {year} {2020})}\BibitemShut {NoStop}%
\bibitem [{\citenamefont {Linowski}\ \emph {et~al.}(2023)\citenamefont {Linowski}, \citenamefont {Schlichtholz}, \citenamefont {Sorelli}, \citenamefont {Gessner}, \citenamefont {Walschaers}, \citenamefont {Treps},\ and\ \citenamefont {Łukasz Rudnicki}}]{Linowski_2023}%
  \BibitemOpen
  \bibfield  {author} {\bibinfo {author} {\bibfnamefont {Tomasz}\ \bibnamefont {Linowski}}, \bibinfo {author} {\bibfnamefont {Konrad}\ \bibnamefont {Schlichtholz}}, \bibinfo {author} {\bibfnamefont {Giacomo}\ \bibnamefont {Sorelli}}, \bibinfo {author} {\bibfnamefont {Manuel}\ \bibnamefont {Gessner}}, \bibinfo {author} {\bibfnamefont {Mattia}\ \bibnamefont {Walschaers}}, \bibinfo {author} {\bibfnamefont {Nicolas}\ \bibnamefont {Treps}}, \ and\ \bibinfo {author} {\bibnamefont {Łukasz Rudnicki}},\ }\bibfield  {title} {\enquote {\bibinfo {title} {Application range of crosstalk-affected spatial demultiplexing for resolving separations between unbalanced sources},}\ }\href {\doibase 10.1088/1367-2630/ad0173} {\bibfield  {journal} {\bibinfo  {journal} {New J. Phys.}\ }\textbf {\bibinfo {volume} {25}},\ \bibinfo {pages} {103050} (\bibinfo {year} {2023})}\BibitemShut {NoStop}%
\bibitem [{\citenamefont {de~Almeida}\ \emph {et~al.}(2021)\citenamefont {de~Almeida}, \citenamefont {Ko\l{}ody\ifmmode~\acute{n}\else \'{n}\fi{}ski}, \citenamefont {Hirche}, \citenamefont {Lewenstein},\ and\ \citenamefont {Skotiniotis}}]{PhysRevA.103.022406}%
  \BibitemOpen
  \bibfield  {author} {\bibinfo {author} {\bibfnamefont {J.~O.}\ \bibnamefont {de~Almeida}}, \bibinfo {author} {\bibfnamefont {J.}~\bibnamefont {Ko\l{}ody\ifmmode~\acute{n}\else \'{n}\fi{}ski}}, \bibinfo {author} {\bibfnamefont {C.}~\bibnamefont {Hirche}}, \bibinfo {author} {\bibfnamefont {M.}~\bibnamefont {Lewenstein}}, \ and\ \bibinfo {author} {\bibfnamefont {M.}~\bibnamefont {Skotiniotis}},\ }\bibfield  {title} {\enquote {\bibinfo {title} {Discrimination and estimation of incoherent sources under misalignment},}\ }\href {\doibase 10.1103/PhysRevA.103.022406} {\bibfield  {journal} {\bibinfo  {journal} {Phys. Rev. A}\ }\textbf {\bibinfo {volume} {103}},\ \bibinfo {pages} {022406} (\bibinfo {year} {2021})}\BibitemShut {NoStop}%
\bibitem [{\citenamefont {Zanforlin}\ \emph {et~al.}(2022)\citenamefont {Zanforlin}, \citenamefont {Lupo}, \citenamefont {Connolly}, \citenamefont {Kok}, \citenamefont {Buller},\ and\ \citenamefont {Huang}}]{zanforlin2022optical}%
  \BibitemOpen
  \bibfield  {author} {\bibinfo {author} {\bibfnamefont {Ugo}\ \bibnamefont {Zanforlin}}, \bibinfo {author} {\bibfnamefont {Cosmo}\ \bibnamefont {Lupo}}, \bibinfo {author} {\bibfnamefont {Peter~WR}\ \bibnamefont {Connolly}}, \bibinfo {author} {\bibfnamefont {Pieter}\ \bibnamefont {Kok}}, \bibinfo {author} {\bibfnamefont {Gerald~S}\ \bibnamefont {Buller}}, \ and\ \bibinfo {author} {\bibfnamefont {Zixin}\ \bibnamefont {Huang}},\ }\bibfield  {title} {\enquote {\bibinfo {title} {Optical quantum super-resolution imaging and hypothesis testing},}\ }\href {\doibase https://www.nature.com/articles/s41534-018-0114-y} {\bibfield  {journal} {\bibinfo  {journal} {Nat. Commun.}\ }\textbf {\bibinfo {volume} {13}},\ \bibinfo {pages} {5373} (\bibinfo {year} {2022})}\BibitemShut {NoStop}%
\bibitem [{\citenamefont {Huang}\ and\ \citenamefont {Lupo}(2021)}]{PhysRevLett.127.130502}%
  \BibitemOpen
  \bibfield  {author} {\bibinfo {author} {\bibfnamefont {Zixin}\ \bibnamefont {Huang}}\ and\ \bibinfo {author} {\bibfnamefont {Cosmo}\ \bibnamefont {Lupo}},\ }\bibfield  {title} {\enquote {\bibinfo {title} {Quantum hypothesis testing for exoplanet detection},}\ }\href {\doibase 10.1103/PhysRevLett.127.130502} {\bibfield  {journal} {\bibinfo  {journal} {Phys. Rev. Lett.}\ }\textbf {\bibinfo {volume} {127}},\ \bibinfo {pages} {130502} (\bibinfo {year} {2021})}\BibitemShut {NoStop}%
\bibitem [{\citenamefont {Lu}\ \emph {et~al.}(2018)\citenamefont {Lu}, \citenamefont {Krovi}, \citenamefont {Nair}, \citenamefont {Guha},\ and\ \citenamefont {Shapiro}}]{lu2018quantum}%
  \BibitemOpen
  \bibfield  {author} {\bibinfo {author} {\bibfnamefont {Xiao-Ming}\ \bibnamefont {Lu}}, \bibinfo {author} {\bibfnamefont {Hari}\ \bibnamefont {Krovi}}, \bibinfo {author} {\bibfnamefont {Ranjith}\ \bibnamefont {Nair}}, \bibinfo {author} {\bibfnamefont {Saikat}\ \bibnamefont {Guha}}, \ and\ \bibinfo {author} {\bibfnamefont {Jeffrey~H}\ \bibnamefont {Shapiro}},\ }\bibfield  {title} {\enquote {\bibinfo {title} {Quantum-optimal detection of one-versus-two incoherent optical sources with arbitrary separation},}\ }\href {https://www.nature.com/articles/s41534-018-0114-y} {\bibfield  {journal} {\bibinfo  {journal} {npj Quantum Inform.}\ }\textbf {\bibinfo {volume} {4}},\ \bibinfo {pages} {64} (\bibinfo {year} {2018})}\BibitemShut {NoStop}%
\bibitem [{\citenamefont {Helstrom}(1973)}]{1055052}%
  \BibitemOpen
  \bibfield  {author} {\bibinfo {author} {\bibfnamefont {C.}~\bibnamefont {Helstrom}},\ }\bibfield  {title} {\enquote {\bibinfo {title} {Resolution of point sources of light as analyzed by quantum detection theory},}\ }\href {\doibase 10.1109/TIT.1973.1055052} {\bibfield  {journal} {\bibinfo  {journal} {IEEE T. Inform. Theory}\ }\textbf {\bibinfo {volume} {19}},\ \bibinfo {pages} {389--398} (\bibinfo {year} {1973})}\BibitemShut {NoStop}%
\bibitem [{\citenamefont {Sacchi}(2005)}]{PhysRevA.71.062340}%
  \BibitemOpen
  \bibfield  {author} {\bibinfo {author} {\bibfnamefont {Massimiliano~F.}\ \bibnamefont {Sacchi}},\ }\bibfield  {title} {\enquote {\bibinfo {title} {Optimal discrimination of quantum operations},}\ }\href {\doibase 10.1103/PhysRevA.71.062340} {\bibfield  {journal} {\bibinfo  {journal} {Phys. Rev. A}\ }\textbf {\bibinfo {volume} {71}},\ \bibinfo {pages} {062340} (\bibinfo {year} {2005})}\BibitemShut {NoStop}%
\bibitem [{\citenamefont {Defienne}\ and\ \citenamefont {Faccio}(2020)}]{PhysRevA.101.063830}%
  \BibitemOpen
  \bibfield  {author} {\bibinfo {author} {\bibfnamefont {Hugo}\ \bibnamefont {Defienne}}\ and\ \bibinfo {author} {\bibfnamefont {Daniele}\ \bibnamefont {Faccio}},\ }\bibfield  {title} {\enquote {\bibinfo {title} {Arbitrary spatial mode sorting in a multimode fiber},}\ }\href {\doibase 10.1103/PhysRevA.101.063830} {\bibfield  {journal} {\bibinfo  {journal} {Phys. Rev. A}\ }\textbf {\bibinfo {volume} {101}},\ \bibinfo {pages} {063830} (\bibinfo {year} {2020})}\BibitemShut {NoStop}%
\end{thebibliography}


\end{document}